\documentclass[aps, twocolumn, 8pt]{revtex4}

\usepackage{graphicx,amssymb}

\def\II{\mathbb I}
\def\NN{\mathbb N}

\newtheorem{Thm}{Theorem}
\newtheorem{Dfn}{Definition}

\begin{document}
\title{Squash Operator and Symmetry}
\author{Toyohiro Tsurumaru}
\affiliation{Mitsubishi Electric Corporation, Information Technology R\&D Center, 5-1-1 Ofuna, Kamakura-shi, Kanagawa, 247-8501 Japan.}

\begin{abstract}
This paper begins with a simple proof of the existence of squash operators compatible with the Bennett-Brassard 1984 (BB84) protocol which suits single-mode as well as multi-mode threshold detectors.
The proof shows that, when a given detector is symmetric under cyclic group $C_{4}$, and a certain observable associated with it has rank two as a matrix, then there always exists a corresponding squash operator.
Next, we go on to investigate whether the above restriction of `rank two' can be eliminated; i.e., is cyclic symmetry alone sufficient to guarantee the existence of a squash operator? The motivation behind this question is that, if this were true, it would imply that one could realize a device-independent and unconditionally secure quantum key distribution protocol. However, the answer turns out to be negative, and moreover, one can instead prove a no-go theorem that any symmetry is, by itself, insufficient to guarantee the existence of a squash operator.
\end{abstract}

\maketitle

Quantum key distribution (QKD) is a technique for distributing information-theoretically-secure secret keys between two parties connected by a quantum channel. The oldest and now defacto standard protocol for QKD is the well-known Bennett-Brassard 1984 (BB84) protocol \cite{BB84}. Several different approaches are known today for proving its unconditional security \cite{SP00,GLLP02,Koashi,HR}; e.g. one based on virtual entanglement distillation protocol (EDP) \cite{SP00,GLLP02} and another based on the complementarity of quantum theory \cite{Koashi}.

The most widely used of these approaches is the one based on EDP, where an actual QKD protocol is converted to an equivalent and virtual EDP performed by Alice and Bob. The conversions must be made so that Alice's and Bob's quantum operations are seen by Eve to remain the same positive operator valued measures (POVM); i.e., Eve's information regarding the secret key bits is not changed by conversion. In the original versions of EDP-based proofs \cite{SP00,GLLP02}, one needed to assume that the actual protocol had access to a perfect single-photon source and photon-number resolving detectors. However, this assumption is invalid for real-world QKD systems, which use attenuated lasers as light sources, and the receiver uses `threshold detectors,' which can discriminate a nonzero photon state from the vacuum, but cannot determine the exact photon number.

In fact, techniques are already known that can fill these gaps. As for light sources, by exploiting decoy states, lasers can be driven to effectively emit single-photon pulses \cite{decoy}. One of the known solutions for detectors \cite{TT08,BML08,KAYN08} is the powerful theoretical tool called `squash operator'. Squash operator is a quantum operation which transforms an incoming $n$-photon state to a qubit state. By incorporating this operator into a conventional type of security proof where Bob has a photon-number discriminating detector, one automatically obtains a new proof that remains valid even if threshold detectors are used. A squash operator was first assumed in the security proof by Gottesman et al. \cite{GLLP02}, however, its existence was only conjectured, no proof was given. For threshold detectors, which are sensitive only to single-mode photon pulses, its existence was proved first by the present author and Tamaki \cite{TT08}, and also independently by Beaudry et al. \cite{BML08}.

The aim of this paper is to investigate how far we can generalize this result from the viewpoint of symmetry constraints imposed on the detector.
In the first half of this paper, we show that when a given set of POVM is symmetric under transformations of cyclic group $C_4$, and the observable $M_z$ related with it has rank two, then there always exists a corresponding squash operator compatible with the BB84 protocol (Theorem 1). An immediate corollary of this theorem is that a squash operator exists, not only for single-mode threshold detectors, but also for multi-mode threshold detectors. 
Next, in the second half of the paper, we tackle the question of whether the above restriction of `rank two' can be eliminated. The answer, however, turns out to be negative. Furthermore, it can be shown that, more generally, no symmetry is sufficient by itself to guarantee the existence of a squash operator (Theorem 2).

{\em Definition of Squash Operator.} ---
In the BB84 protocol, Alice and Bob use two different bases, $r$, for their measurements, interchangeably. They are usually denoted as the $z$ and the $x$ basis ($r=z,x$) because they are related to qubit measurements of the Pauli matrices $\sigma_z, \sigma_x$. Similarly, the notation of $r=+,\times$ bases is used to indicate the directions of photon polarization. In what follows, we stick to the notation of $r=z,x$ for the sake of simplicity.

We denote the Hilbert space of the receiver's incoming states as ${\cal H}_B$. In this space, there are two sets of POVM elements, $M_{(r,b)}$, corresponding to basis $r=z,x$ and the output bit $b=0,1$. We also define observables $M_r:=M_{(r,0)}-M_{(r,1)}$ for later convenience. For example, if a receiver measures state $\rho_B\in{\cal H}_B$ using the $x$ basis, he observes output bit $b=0$ with the probability $p_{(x,0)}={\rm Tr}\left(\rho_B M_{(x,0)}\right)$. We also assume that the measurements are complete for each basis; that is,
\begin{equation}
M_{(r,0)}+M_{(r,1)}=\II_B
\label{eq:assumption1}
\end{equation}
holds for $r=z,x$, where $\II_B$ is the identity operator of ${\cal H}_B$.

Squash operator $F$ is a completely positive trace-preserving (CPTP) map with the following properties \footnote{Squash operation is called `squashing model' in \cite{BML08}.}.
$F$ maps states in ${\cal H}_B$ to those in qubit space ${\cal H}_C$, and, when $F$ is followed by the $z$ or the $x$ measurement in ${\cal H}_C$, it reproduces $M_r$ of the actual measurement device. That is, for an arbitrary mixed state $\rho_B\in{\cal H}_B$, it satisfies
\begin{equation}
{\rm Tr}\left(F(\rho_B)\sigma_r\right)={\rm Tr}\left(\rho_B M_r\right)\ {\rm for}\ r=z,x
\label{eq:cond_squash}
\end{equation}
with $\sigma_r$ being the Pauli operators.
A convenient way of describing $F$ is to use an operator sum representation with a set of Kraus operators $F_c$ (see, e.g., \cite{Nielsen}.)
In this notation, the trace-preserving condition of $F$ takes the form $\sum_cF^\dagger_cF_c=\II_B$. Complete positiveness is guaranteed as long as $F$ is expressed as $F(\rho_B)=\sum_cF_c\rho_B F^\dagger_c$.
This notation has the additional merit that the Hermitian conjugate, $F^\dagger$, of $F$ can be expressed in simple form as $F^\dagger(\rho_C)=\sum F_c^\dagger\rho_CF_c$ with $\rho_c$ being an arbitrary state in ${\cal H}_C$.
By using these relations, the definition of squash operator $F$ for $M_r$ given in (\ref{eq:cond_squash}) can be equivalently stated as the following two conditions for Kraus operator $F_c$,
\begin{eqnarray}
M_r&=&\sum_cF_c^\dagger\sigma_rF_c,\\
\II_B&=&\sum_cF^\dagger_cF_c.
\end{eqnarray}

{\em Cyclically Symmetric POVM for the BB84 protocol.} ---
In the first half of this paper, we show that $F$ actually exists for multi-mode threshold detectors as well.
Against this goal, we generalize the problem slightly by taking up finite group $C_4$, i.e., a cyclic group of order $4$, and consider POVM elements $M_{(r,b)}$ which are symmetric under its transformations (for details of $C_4$ group, see, e.g., \cite{Serre}.)
The $C_4$-symmetry of $M_{(r,b)}$ is stated rigorously as follows.
\begin{Dfn}
A set of POVM elements $\{M_{(r,b)}\}$ of BB84 type is $C_4$-symmetric, if there exists unitary operator $U$ satisfying $U^{4k}=\II_B$ with $k\in\NN$, and it transforms them as follows
\begin{eqnarray}
UM_{(z,b)}U^\dagger&=&M_{(x,b)},\label{eq:rel1}\\
U^2M_{(r,b)}U^{\dagger2}&=&M_{(r,1-b)}.\label{eq:rel2}
\end{eqnarray}
\end{Dfn}
Intuitively, operator $U$ corresponds to rotating a detector spatially by 45 degrees, when polarization encoding is used. It can be better seen if we newly define operators $L_0,\dots,L_3$ as $L_{2b}=M_{(z,b)}$ and $L_{2b+1}=M_{(x,b)}$ for $b=0,1$. The relations (\ref{eq:rel1}) and (\ref{eq:rel2}) can thus be rewritten as $UL_{c}U^\dagger=L_{c+1}$, where modulo 4 is assumed in the summation of index $c$. Note here that, with $U$ being a 45-degree rotation, we have $U^8=\II_B$ instead of $U^4=\II_B$. This example demonstrates why we needed to consider cases of $k>1$ in Definition 1.

\begin{Thm}
If a given set of POVM elements $\{M_{(r,b)}\}$ of BB84 type is $C_4$-symmetric, and the rank of the corresponding observable $M_z$ (or equivalently, $M_x$) as a matrix is two, there always exists a corresponding squash operator compatible with the BB84 protocol. 
\end{Thm}
Here, it should be noted that the restriction of `rank two' does not necessarily mean that the Hilbert space ${\cal H}_B$ is a qubit space, as illustrated by the following example.

An important example of $C_4$-symmetric POVMs is the threshold detector.
In this paragraph, following \cite{TT08,BML08}, we concentrate on photon detection modules consisting of two photon threshold detectors, each of which corresponds to output bits $b=0,1$;
we call such photon detection units simply `threshold detectors' with a slight abuse of the terminology.
We also assume that, when both detectors click coincidently (double-click events), the detection system outputs a random bit as its output, $b$.
However, we differ from \cite{TT08,BML08} in that we do not restrict ourselves to a single mode, but assume that an incoming light pulse may have $m\ge1$ modes of propagation; we label each of them by index $i$. We also denote the number of photons in mode $i$ as $n_i$, and let $N=(n_1,n_2,\dots,n_m)$. Clearly, any threshold detector is block diagonalized with respect to the photon number configuration $N$, and there is no loss of generality in considering each of the blocks individually when analyzing security. For each such section, $N$, the observables $M_r$ can be written as a matrix with rank two
\begin{equation}
M_r=|N;r,0\rangle\langle N;r,0|
-|N;r,1\rangle\langle N;r,1|
\label{eq:def_M_r}
\end{equation}
for $r\in\left\{z,x \right\}$, where
\begin{equation}
|N;r,b\rangle:=A_N(a_{1rb}^\dagger)^{n_1}(a_{2rb}^\dagger)^{n_2}\cdots(a_{mrb}^\dagger)^{n_m}|0\rangle.
\label{eq:def_Nrb}
\end{equation}
Note here that $M_r$ has rank two because double-click events are replaced by a random bit in our model, and thus all states besides $|N;r,0\rangle$ and $|N;r,1\rangle$ are cancelled in the subtraction $M_r=M_{(r,0)}-M_{(r,1)}$ (a similar argument can be found in \cite{KAYN08}.)
Coefficient $A_N$ in (\ref{eq:def_Nrb}) is the normalization constant for state $|N;r,b\rangle$, and 
$a^\dagger_{irb}$ are the creation operators for photons propagating in mode $i$, having bit value $b$ of basis $r$.
In accordance with the usual notations of Pauli matrices, creation operators $a^\dagger_{irb}$ for two bases $r=z,x$ are related as $a^\dagger_{ixb}=\frac1{\sqrt2}\left(a^\dagger_{iz0}+(-1)^b a^\dagger_{iz1}\right)$.
The single-mode threshold detectors discussed in \cite{TT08,BML08} correspond to the special case of $m=1$.
$C_4$-symmetry can be shown by using an explicit form of the transforming operator $U_N$,
\[
U_N=\exp\left[\frac{i}2\sum_i\left(a^\dagger_{iy0}a_{iy0}-a^\dagger_{iy1}a_{iy1}\right)\right].
\]
The creation operator along the $y$-axis appearing in the above equation is defined as $a^\dagger_{iyb}=(a^\dagger_{iz0}+i(-1)^b a^\dagger_{iz1})/\sqrt2$.

From these facts, and also from Theorem 1, it immediately follows that a squash operator exists, not only for single-mode threshold detectors, but also for multi-mode threshold detectors.

{\em Proof of Theorem 1.}
Here we give only the proof for $k=1$, since all other cases ($k\ge2$) can be shown by exactly the same argument. As can be seen from $U^2M_zU^{\dagger2}=-M_z$, for each normalized eigenstate $|v\rangle$ of $M_z$ with an eigenvalue $0<\lambda\le1$, there always exists another eigenstate $U^2|v\rangle$ having a different eigenvalue $-\lambda$, and thus is orthogonal
\begin{equation}
\langle v|U^2|v\rangle=0.
\label{eq:orthogonality}
\end{equation}
From this, and since the rank of $M_z$ is two, it follows that $M_z$ takes the form
\begin{equation}
M_z=\lambda\left(|v\rangle\langle v|-U^2|v\rangle\langle v|U^{\dagger2}\right).
\label{eq:N_expanded}
\end{equation}
By using a basis that diagonalizes $U$, we can always decompose $|v\rangle$ as
\begin{equation}
|v\rangle=\sum_{c=0}^3\mu_c\left|v_c\right\rangle,
\label{eq:v_expanded}
\end{equation}
with $U\left|v_c\right\rangle=i^c\left|v_c\right\rangle$. Then, from (\ref{eq:orthogonality}), we see that coefficients $\mu_i$ satisfy
\begin{equation}
|\mu_0|^2+|\mu_2|^2=|\mu_1|^2+|\mu_3|^2=\frac12.
\label{eq:mu_relation}
\end{equation}

We now define a completely positive, but not necessarily trace-preserving map, $F$, with a set of Kraus operators $F_0,F_1,\dots,F_3$ which take the form
\begin{equation}
F_c=
\sqrt{2\lambda}\left(\mu_{c+1}|0_y\rangle\langle v_c|+\mu_c|1_y\rangle\langle v_{c+1}|\right)
\label{eq:define_F_c}
\end{equation}
if $\mu_c\mu_{c+1}^*\ne0$, otherwise $F_c=0$. In (\ref{eq:define_F_c}), modulo 4 is assumed for the summations of index $c$. From the linearity of $F^\dagger$, we obtain the following relation
\begin{eqnarray}
\lefteqn{F^\dagger(\sigma_z+i\sigma_x)=\sum_{c=0}^3F^\dagger_c(\sigma_z+i\sigma_x)F_c}\nonumber\\
&=&2\sum_{c=0}^3F^\dagger_c|0_y\rangle\langle1_y|F_c
=4\lambda\sum_{c=0}^3\mu_c\mu_{c+1}^*\left|v_c\right\rangle\left\langle v_{c+1}\right|\nonumber\\
&=&\lambda\sum_{c=0}^3i^cU^c|v\rangle\langle v|U^{\dagger c}=M_z+iM_x.
\label{eq:F_relation}
\end{eqnarray}
Similarly, from eqs. (\ref{eq:mu_relation}) and (\ref{eq:define_F_c}), we have $F^\dagger(\II_C)=\sum_{c=0}^3F_c^\dagger F_c\le \II_B$. Furthermore, $F$ can be modified such that it satisfies the trace-preserving condition (\ref{eq:rel2}), and also maintains relation $F^\dagger(\sigma_z+i\sigma_x)=M_z+iM_x$, obtained in (\ref{eq:F_relation}). This can be done by introducing extra Kraus operators $F_c$, $c>3$, having the form $F_c=|b_y\rangle\langle\psi_c|$ with $b=0$ or $1$.

That the CPTP map $F$ thus obtained also satisfies (\ref{eq:rel1}) for $r=z$ can be shown as $F^\dagger(\sigma_z)=\frac12F^\dagger\left((\sigma_z+i\sigma_x)+{\rm H.c.}\right)=\frac12(M_z+iM_x)+{\rm H.c.}=M_z$, where `H.c.' denotes the Hermitian conjugate. The other relation for $r=x$ can be shown similarly.
(End of proof.)

{\it Does Symmetry Imply the Existence of Squash Operators?} ---
A natural question that arises here is: Can we eliminate the restriction of `rank two' appearing in Theorem 1? In other words, is cyclic symmetry $C_4$ alone sufficient to guarantee the existence of a squash operator? Or more generally, is there any types of symmetry that is strong enough to ensure its existence?
In the remaining half of this paper, we shall investigate this possibility.
This question is interesting because if this were actually the case, we would need no knowledge about microscopic structures of a detector in order to ensure the existence of its squash operator. In other words, we would succeed in proving the unconditional security of QKD in a device-independent way (for security against collective attacks, see \cite{Acin07}).

Indeed, $C_4$-symmetry is already realized in most conventional BB84 systems (c.f. the paragraph below Definition 1). For example, when polarization encoding is used, bases $z,x$ can be switched by rotating the detector by 45 degrees. In addition, the receiver may interchange the assignment of two detectors to output bit $b=0,1$ randomly, by rotating them by 90 degrees and flipping $b$. This may be done, in order to cancel the mismatch between two detectors in terms of quantum bit error rate. These two types of rotation generate a $C_4$ group.

Moreover, for some QKD protocols, no knowledge about microscopic structures of any components {\it besides detectors} are needed to prove security. For example, consider the Bennett-Brassard-Mermin 1992 protocol \cite{BBM92}, where an untrusted third party prepares an entangled state. It is clear that if symmetry could actually imply the existence of squash operators, we would be able to prove the security of a QKD system without knowing anything of the microscopic structure of the devices, only the macroscopic operations by Alice and Bob.

However, as we shall show below, that is not actually the case. The fact is that we can prove a no-go theorem that denies all such possibilities.
In order to discuss this point rigorously, we define general symmetries of POVM below, and then present a theorem.
\begin{Dfn}
A set of POVM elements $\{M_{(r,b)}\}$ of BB84 type is symmetric under finite group $G$, if they transform under $G$ as
\[
V(g)M_{(r,b)}V^\dagger(g)=M_{g(r,b)}\ {\rm for}\ g\in G,
\]
with $V(g)$ being a unitary representation of group $G$.
Here, map $g:(r,b)\mapsto(r',b')$ determines how each POVM element $M_{(r,b)}$ is transformed into another element by $g\in G$.
\end{Dfn}

\begin{Thm}
No symmetry is sufficient by itself to guarantee the existence of a squash operator.
That is, one cannot prove a theorem that states that `For an arbitrary $G$-symmetric set of POVM, there always exists a squash operator compatible with the BB84 protocol.'
\end{Thm}

We present below a proof of this theorem. The basic strategy here is to show that, if the type of theorems as quoted in Theorem 2 holds, it can be used to show the improbable proposition that any arbitrary operator, whether symmetric or asymmetric, possesses a squash operator.

{\em Proof of Theorem 2.}
For an arbitrary set of operators $M_{(r,b)}$ of BB84 type, which may not be symmetric, one can always define other $G$-symmetric operators $\tilde{M}_{(r,b)}$ as
\[
\tilde{M}_{(r,b)}:=\sum_{g\in G}M_{g^{-1}(r,b)}\otimes |g\rangle\langle g|
\]
in ${\cal H}_B \otimes{\cal H}_D$. Here, ${\cal H}_D$ is an ancilla space which is spanned by orthonormal basis $\left\{|g\rangle\right\}_{g\in G}$, that is, a set of orthonormal states $|g\rangle$ labeled by all elements $g\in G$.
$\tilde{M}$ is $G$-symmetric under unitary transformation $\tilde{V}$ as defined by $\tilde{V}(g):={\rm id}_B\otimes R_D(g)$ with $R_D$ being the regular representation of $G$ defined by $R_D(g)|h\rangle_D=|gh\rangle_D$ (see, e.g., \cite{Serre}).
Hence if one could prove the type of theorems quoted in Theorem 2, it would readily follow that there is a squash operator for $\tilde{M}_{(r,b)}$.

On the contrary, however, once such a $\tilde{F}$ is obtained, one can also construct squash operator $F$ for the original operators $M_{(r,b)}$.
In order to see this, note that we have for an arbitrary $\rho_B\in{\cal H}_B$
\begin{eqnarray*}
{\rm Tr}\left[M_r\rho_B\right]&=&{\rm Tr}\left[\tilde{M}_r\left(\rho_B\otimes|e\rangle\langle e|\right)\right]\\
&=&{\rm Tr}\left[\sigma_r\tilde{F}(\rho_B\otimes|e\rangle\langle e|)\right].
\end{eqnarray*}
with $e$ being the identity element of $G$.
Thus, applying $\tilde{F}$ on $\rho_B\otimes|e\rangle\langle e|$ serves as a correct squash operator for $\rho_B$.
This result shows that any POVM $M_{(r,b)}$ of BB84 type, whether symmetric or not, possesses a squash operator. However, this leads to a contradiction because there exists the counterexample of POVM $M_0$ defined by
\[
M_{0z}=M_{0x}=\sigma_z
\]
which has no squash operator.

The fact that $M_0$ does not possess any squash operator can be shown, e.g, by the same argument as Beaudry, Moroder, and L\"{u}tkenhaus used for the six-state protocol \cite{BML08}, but it can alternatively be shown by the following simple argument.
Consider the situation where Alice and Bob perform the BBM92 protocol using $M_0$ as their detectors, and, as the entanglement source, Eve provides states $|\psi_b\rangle:=|b_z\rangle_A\otimes|b_z\rangle_B$ with a bit $b\in\{0,1\}$ of her choice. In this setup, clearly, all sifted key bits $b$ are known to Eve, and thus Alice and Bob will never succeed in sharing secret keys. Hence, the existence of squash operator $F$ for $M_0$ would lead to a contradiction, since $F$ could be used to prove the unconditional security of this system, with the quantum bit error rate measured by Alice and Bob being exactly zero.
(End of proof.)

{\em Summary.} ---
In this paper, we first showed that, if a given detector is $C_4$-symmetric, and the observable $M_z$ associated with it has rank two, then there always exists a corresponding squash operator compatible with the BB84 protocol (Theorem 1). By using this result, we then proved that squash operators exist, not only for single-mode threshold detectors, but also for multi-mode threshold detectors.

Next, we took up the question of whether this result can be generalized to symmetric detectors with arbitrary ranks, as an attempt toward the realization of device-independent and unconditionally secure QKD protocols. However, it turned out that the fact is quite opposite. That is, we have succeed in proving that, no matter what symmetry one imposes on the detectors, the symmetry is never sufficient, by itself, to guarantee the existence of a corresponding squash operator (Theorem 2).

{\em Acknowledgments} ---
The author would like to thank M. Koashi and K. Tamaki for their valuable comments.
This work was supported by the National Institute of Information and Communications Technology (NICT), Japan.

\end{document}